
\font\twelvebf=cmbx10 scaled \magstep1
\hsize 15.0truecm \hoffset 1.0 truecm
\vsize 22.0truecm
\nopagenumbers
\headline={\ifnum \pageno=1 \hfil \else\hss\tenrm\folio\hss\fi}
\pageno=1

\hfill IUHET 302

\hfill IU/NTC 95--06

\hfill May 26, 1995
\bigskip
\centerline{\twelvebf New formulas relating the masses
of some baryons and mesons}

\medskip
\bigskip

\centerline{D. B. Lichtenberg and  R. Roncaglia}

\centerline{Physics Department, Indiana University,
Bloomington, IN 47405, USA}

 \vskip 1 cm

Sum rules relating the masses of ground-state baryons and
mesons are obtained
in a constituent quark model. The interaction is
assumed to be independent of quark spins except for a
spin-dependent part that can be
treated as a perturbation.
Where data are available, the sum rules agree with
experiment to better than 1\%.

\bigskip
\noindent PACS numbers: 12.10.Kt, 12.40.Yx

\bigskip
\medskip \bigskip \bigskip

In this work we consider meson and baryon masses using a
Hamiltonian formalism within the framework of the
constituent quark model. We do not assume any particular
functional form for the spin-independent part of the
Hamiltonian. However, we treat the spin-dependent part
as a perturbation, and we confine ourselves to
ground-state hadrons, so that the spin dependence arises
from the spin-spin or colormagnetic term. With these
modest inputs plus an additional approximation (to be
discussed later), we are able to obtain three sum rules
relating the masses of certain baryons and mesons. Two
of these sum rules agree with experiment to better than
1\% and the third remains to be tested.

In the constituent quark model, the mass of a hadron
can be written as the sum of the masses of its constituent
quarks plus an interaction energy. The mass $M_M(ij)$
of a meson composed of a quark $q_i$ and an antiquark
$\bar q_j$ with masses $m_i$ and $m_j$ respectively is
given by
$$M_M(ij)= m_i +m_j + E(ij), \eqno(1)$$
where $E(ij)$ is the interaction energy.
Similarly, the mass $M_B(ijk)$ of a baryon containing
quarks with masses $m_i$, $m_j$, and $m_k$ is
$$M_B(ijk)= m_i + m_j +m_k + E(ijk),  \eqno(2)$$
where $E(ijk)$ is the interaction energy.  We
assume that the interaction energies $E(ij)$ and $E(ijk)$,
and consequently the masses $M_M(ij)$ and $M_B(ijk)$ are
calculable in principle from a Hamiltonian without
spin-dependent forces. We also assume that the mass of a
constituent quark can be assigned a unique value,
independent of the hadron in which it is bound, but we do
not have to specify the quark masses to obtain our results.

It follows from Eqs.\ (1) and (2) that
$$M_B(123) -M_B(453) - M_M(12) +M_M(45)  =
\delta(12345) ,  \eqno(3)$$
where $\delta(12345)$ is defined by
$$\delta(12345) =E(123)- E(453) - E(12) +E(45). \eqno(4)$$
We immediately see from Eq.\ (3) that if $\delta(12345)$
is zero, we have the sum rule
$$M_B(123) - M_M(12) =  M_B(453)  -M_M(45). \eqno(5)$$

We next come to the important question of the circumstances
under which Eq.\ (5) is most likely to hold to a good
approximation. It has been previously found [1,2] that
the ground-state vector meson and spin 3/2 baryon
interaction energies are smooth, monotonically decreasing
functions of the generalized reduced mass $\mu(12...)$ of
the system, where $\mu(12...)$ is defined by
$$1/\mu(12...) = \sum_i 1/m_i. \eqno(6)$$
We have looked at the energies $E(ij)$ and $E(ijk)$
appropriately averaged over the spins of ground-state
hadrons of different spins and the same quark content. We
have verified numerically that, for reasonable values of
the quark constituent masses, these energies are also smooth
functions of $\mu$.

We next observe that $E(ij)$ and $E(ijk)$, considered
as functions of $\mu(ij)$ and $\mu(ijk)$ respectively,
depend most sensitively on the mass of the lightest
quark. We therefore tend to have a small change in
$\mu(ij)$ and $\mu(ijk)$ if we confine ourselves to
the case in which initially all the quarks in a hadron
are $u$ or $d$ quarks (denoted by $n$), and only one of
the quarks (say the $j$ quark) is allowed to change from
an $n$ quark to a heavier $s$, $c$, or $b$ quark (denoted
by $h$).
Because the $E(ij)$ and $E(ijk)$ are smooth functions
of $\mu(ij)$ and $\mu(ijk)$ respectively, small
changes in the reduced masses will lead to small changes
in the interaction energies. Consequently, if
$\mu(453)$ is close to $\mu(123)$ and if $\mu(45)$ is
close to $\mu(12)$, the value of $\delta(12345)$, given
in Eq.\ (4), will be small and we shall  neglect it.
We insure this to be the case by confining ourselves
to hadrons containing at most one quark which is not
a light $n$ quark; specifically, we assume
$$\delta(nnnnq) =0, \eqno(7)$$
where $q$ is any quark.
Then, from Eq.\ (5)  we have  the sum rule
$$M_B(nnn) - M_M(nn) =  M_B(nhn)  -M_M(nh). \eqno(8)$$

We now consider the effect of spin-dependent forces.
We let the spin-dependent colormagnetic interaction be
a perturbation of the form [1]
$$V_{cm}= -\sum_{i<j}
 \sigma_i \cdot \sigma_j \lambda_i \cdot
\lambda_j f(r_{ij})/(m_i m_j), \eqno(9)$$
where the $\sigma_i$ are Pauli spin matrices,
the $\lambda_i$ are Gell-Mann matrices, and $f(r_{ij})$ is
a positive-definite operator which depends
on the separation $r_{ij}$  between quarks $i$ and $j$.  We
do not need to specify the functional form of $f(r_{ij})$.
Then it is straightforward to take hadron spin averages
[3]. We obtain three sum rules from Eq.\ (8):
$$(N+\Delta)/2-(3\rho +\pi)/4 =
(2\Sigma^* + \Sigma + \Lambda)/4
-(3K^* + K)/4 \eqno(10)$$
if $h=s$,
$$(N+\Delta)/2-(3\rho +\pi)/4 =
(2\Sigma_c^* + \Sigma_c + \Lambda_c)/4
-(3D^* + D)/4  \eqno(11)$$
if $h=c$, and
$$(N+\Delta)/2-(3\rho +\pi)/4 =
(2\Sigma_b^* + \Sigma_b + \Lambda_b)/4
-(3B^* + B)/4  \eqno(12)$$
if $h=b$,
where the symbol for a hadron denotes its mass.

We can test the first of these sum rules with data from
main tables of the Particle Data Group [4]. We obtain
that the left hand side of Eq.\ (10) is 474 Mev, while
the right hand side is 473 MeV, giving excellent agreement
with experiment. Turning to Eq.\ (11), we find that
all masses except that of the $\Sigma_c^*$ are well
known, but the $\Sigma_c^*$ is given as a poorly
established state of mass $2530\pm 10$ MeV. We therefore
use Eq.\ (11) to predict the mass of the $\Sigma_c^*$,
obtaining
$$\Sigma_c^*= 2523\ {\rm MeV}. \eqno(13)$$
The agreement between our prediction and the value
given by the Particle Data Group is sufficiently close
to lend theoretical support to the experimental
evidence [5] that the $\Sigma_c^*$ baryon has
indeed been found. At the present time, the sum
rule given in Eq.\ (12) cannot be tested because the
mass of the $\Lambda_b$ is known only with an error
of 50 MeV and the $\Sigma_b$ and $\Sigma_b^*$ have not
yet been observed. However, we hope that
Eq.\ (12) can be tested in the not-too-distant future.

In conclusion, we have obtained three sum rules relating
meson and baryon masses. Two of these formulas agree with
existing experimental data, and the third should be testable
by future experiments. Although we have obtained our results
within the framework of the constituent quark model and the
Hamiltonian formalism, we have not had to give  explicitly
either the Hamiltonian or the quark masses.

This work was supported in part by the U. S. Department
of Energy and in part by the U. S. National Science
Foundation.

\vskip 2cm

References
\bigskip

\item{[1]} R. Roncaglia, A. Dzierba, D.B. Lichtenberg,
and E. Predazzi, Phys. Rev. D 51, 1248 (1995).

\item{[2]} R. Roncaglia, D.B. Lichtenberg,
and E. Predazzi, Indiana Univ. report No.  IUHET 293
(1995), Phys.\ Rev.\ D (to be published).

\item{[3]} M. Anselmino, D. B. Lichtenberg, and E.
Predazzi, Z. Phys. C {\bf 48}, 605 (1990).

\item{[4]} Particle Data Group: L. Montanet et al.,
Phys. Rev. D {\bf 50}, 1173 (1994).

\item{[5]} V. V. Ammosov et al., Pis'ma
Zh.\ Eksp.\ Teor.\ Fiz.\  {\bf 58}, 241 (1993)
[JETP Lett.\ {\bf 58}, 247 (1993)].

\bye